\documentclass[preprint,aip,graphicx]{revtex4}
\usepackage{graphicx}
\usepackage{dcolumn}
\usepackage{bm}
\usepackage{amssymb}
\usepackage{color}
\usepackage{caption}
\usepackage{epstopdf}
\begin{document}
\title{Predicting protein dynamics from structural ensembles.}
\author{J. Copperman}
\affiliation{Department of Physics, University of Oregon, Eugene, Oregon 97403}
\author{M.G. Guenza\footnote{Author to whom correspondence should be addressed. Electronic mail: mguenza@uoregon.edu}}
\affiliation{Department of Chemistry and Institute of Theoretical Science, University of Oregon, Eugene, Oregon 97403}
\date{\today}

\begin{abstract}
The biological properties of proteins are uniquely determined by their structure and dynamics. A 
protein in solution 
\textcolor{black}{populates a structural ensemble of metastable configurations 
around the global fold.} From 
\textcolor{black}{overall} rotation to local fluctuations, the dynamics of proteins can cover several orders of magnitude in time scales. We propose a simulation-free coarse-grained approach which utilizes knowledge of the important metastable folded states of the protein to predict the protein dynamics. This approach is based upon the Langevin Equation for Protein Dynamics (LE4PD), a Langevin formalism in the coordinates of the protein backbone. The linear modes of this Langevin formalism organize the fluctuations of the protein, so that more extended dynamical cooperativity relates to increasing energy barriers to mode diffusion. The accuracy of the LE4PD is verified by analyzing the predicted dynamics across a set of seven different proteins for which both relaxation data and NMR solution structures are available. Using experimental NMR conformers as the input structural ensembles, LE4PD predicts quantitatively accurate results, with correlation coefficient $\rho=.93$ to NMR backbone relaxation measurements for the seven proteins. The NMR solution structure derived ensemble and predicted dynamical relaxation is compared with molecular dynamics simulation-derived structural ensembles and LE4PD predictions, and are consistent in the timescale of the simulations. The use of the experimental NMR conformers frees the approach from computationally demanding simulations.
\end{abstract}



\maketitle
\section{Introduction}
\textcolor{black}{The evolved amino acid sequence of a native protein encodes its folded structure and inherent dynamical properties in aqueous solution.}\cite{pettitt1990advances,frauenfelder1991energy,lewandowski2015direct} The latter determines the dynamics of specific residues in a protein primary sequence, which are active participants in the pathways of the biological function. Biologically active segments are often mobile and adaptable to assume a proper configuration when binding to a reaction partner. The multiple configurational states that an active segment may populate are not randomly selected: configurations with minimal energy are connected by energy barriers, and as such are thermally activated, enabling emerging regions of high mobility, which can behave like ``switches" along the binding pathway.\cite{lewandowski2015direct} 

Different experiments and computational models exist to probe the dynamical processes of proteins, spanning the femtosecond regime of bond and angle vibrational modes to the millisecond and longer time regimes of folding and enzymatic kinetics. Important information in the picosecond to tens of nanosecond regime can be collected through NMR relaxation experiments, such as $T_1$, $T_2$ and $NOE$, however their interpretation is model dependent. Atomistic Molecular Dynamic (MD) simulations can provide a realistic dynamical model, but for most proteins of interest sufficient sampling to obtain converged dynamical correlations is prohibitively costly, and a theoretical approach is needed.

The theory we present here is the Langevin Equation for Protein Dynamics (LE4PD), which provides a coarse-grained but still physically realistic representation of biological macromolecules at the lengthscale of a single amino acid and larger. The LE4PD theory describes the amino acid dynamics quantitatively, as the theory contains information about the extent of the intramolecular energy barriers, specific amino acid friction coefficient, semiflexibility, degree of hydrophobicity, as well as hydrodynamics. The LE4PD accurately predicts the sequence-dependent dynamics starting from the ensemble of metastable structural configurations around the folded state measured by NMR, or from MD simulations.

\textcolor{black}{The LE4PD model is unique in that it is a minimal dynamical model which projects the local and global diffusive dynamics of proteins from the protein structural ensemble with no adjustable parameters. This is possible because it is a coarse-grained yet microscopic model whose parameters are set directly from the microscopic physical system. This is in contrast to most methods constructed to model protein dynamics which rely upon site-specific adjustable parameters, such as the model-free formalism of Lipari and Szabo.\cite{lipari1982model} Other methods attempt to define the internal diffusion of proteins as fractional Brownian processes,\cite{kneller2004fractional} which is a more accurate description of the general nature of the internal motion of proteins but is not predictive in nature. Nodet, Abergel, and Bodenhausen have modeled the dynamics of proteins as a coupled network of rotators under the assumption of a single conformational minima and small displacements.\cite{abergel2005predicting} This approach, which attempts to predict fluctuations and dynamics from a single protein structure, is not directly comparable to the LE4PD model we present here, where we model the dynamics and take the structural ensemble from experiment or by sampling an underlying atomistic model via MD simulation. Like other elastic network models\cite{go1983dynamics,bahar1997direct,atilgan2001anisotropy,yang2007well,yang2009protein} the coupled rotator model is capable of capturing the local variation in flexibility along the protein chain with no site-specific adjustable parameters, but because it begins from an empirical network description it requires a large amount of parameterization and specification of an overall rotational diffusion time $\tau_0$, a scaling factor $k_0$, a cut-off distance $R_c$, and a characteristic internal diffusion time $t_D$. The model is explicitly limited to small displacements around a single conformational minima, and relaxation times centered upon a short characteristic internal diffusion time of $\sim300ps$. In contrast, the LE4PD model is capable of simultaneously describing the global rotational diffusion, as well as local motion spanning the picosecond to many nanosecond and microsecond regimes. In particular, the long-time, highly correlated, large-amplitude dynamical motion of proteins is of great biological interest.}

Input to the LE4PD is an ensemble of structural configurations, which has to be representative of the distribution of folded states of the protein. While proteins sample a very large  \textcolor{black}{$3 N$-dimensional} configurational space, with $N$ the number of independent sites comprising the protein, at the bottom of the funnel-like energy landscape the conformational diversity is much smaller.\cite{bryngelson1995funnels,onuchic1997theory,dill1997levinthal} A common paradigm is that the important internal fluctuations of a folded protein span a limited number of specific structures,\cite{monod1965nature,levinthal1968there} and these can be well sampled  experimentally by NMR.\cite{lange2008recognition} If that is the case, NMR conformer ensembles should provide a structural ensemble consistent with well-sampled MD simulations, and the LE4PD coupled with structural NMR should provide predictions of the protein dynamics without need of performing lengthy computer simulations. \textcolor{black}{In practice, NMR solution structures encode a structural diversity that is due to a combination of thermal fluctuations and a possible lack of complete experimental information. The LE4PD method provides the ability to test the capability of an input structural ensemble to produce experimentally determined dynamical measurements such as site-specific NMR relaxation.}

\textcolor{black}{The diffusive mode solution of the LE4PD organizes the configurational landscape, defining fluctuations on a set of well-defined length and timescales encompassing the relative motion between neighboring $\alpha$-carbons to the global rotations of the structure as a whole.}\cite{caballero2007theory,copperman2014coarse-grained} In the diffusive mode description the LE4PD identifies the regions of local flexibility and cooperative motion of the residues inside a protein. As an example we project the MD trajectory onto the diffusive modes of the HIV protease monomer, and obtain a free energy landscape barrier height distribution which scales with mode cooperativity. Using the scaling form for this barrier height distribution, which appears to be a general feature of protein dynamics, leads to accurate dynamical timescales in the simulation-free conformer-based LE4PD model.

\textcolor{black}{Mode-based descriptions are extremely useful in computational approaches to protein dynamics.\cite{amadei1993essential,berendsen2000collective} Analysis of the free energy landscape in covariance modes have been used to describe the folding of small proteins.\cite{maisuradze2009principal} The covariance matrix of the spatial functions of the nuclear spin interactions from MD simulation have been used to calculate NMR relaxation, as fit to the trajectory correlation times and experimental values.\cite{prompers2001reorientational} The characteristic difference between the LE4PD approach and these other approaches is that we study the modes of an appropriate equation of motion, and as such are associated directly with the timescale and pathway of a quasi-independent structural relaxation process. Other mode-based approaches are based upon studying the abstract covariance modes of a set of variables, and as such any time-dependence in these modes comes purely from a fit to the simulation trajectory.}

The dynamical predictions of the LE4PD model starting from an ensemble of structures generated from experimentally determined NMR conformers are compared with a second ensemble of structures generated in the course of an MD simulation in the timescale of $50-150$ nanoseconds. To validate the accuracy of the theoretical predictions of the dynamics using the LE4PD approach we test its predictions against experimental data of NMR relaxation for seven different proteins and 1876 site-specific NMR relaxation measurements. Using either the MD generated or NMR solution structure ensembles we obtain quantitatively self-consistent predictions, with similar overall correlation of $\rho_{MD}=.95$ and $\rho_{NMR}=.93$. We find that, in general, the MD-generated ensembles provide through the LE4PD a closer agreement with experimental data than the LE4PD informed by NMR ensembles, with 17\% lower relative error.

\section{Theoretical Approach: the Langevin Equation for Protein Dynamics}
\label{theory}
In the LE4PD equation the dynamics of the protein is described as a diffusive motion across the configurational landscape,\cite{caballero2007theory,copperman2014coarse-grained,perico1995protein} consistent with an optimized Rouse-Zimm theory of the dynamics of macromolecules in solution.\cite{doi1986theory,perico1985viscoelastic} Proteins are anisotropic in shape and have a hydrophobic core which is only partially exposed to solvent, with this effect  depending on the position of each amino acid in the protein. The LE4PD includes both rotational anisotropy and the hydrophobic core, which are features characteristic of biological macromolecules but are uncommon in synthetic polymers in solution. Local energy barriers in the interior of the protein are important to properly define its dynamics and are explicitly taken into account in the LE4PD method.

The Langevin equation formalism is derived  starting from the Liouville equation for the conservation of probability density in the phase space of the full atomistic system of the protein and solvent, and using projection operators to obtain an equation of motion for the chosen sites.\cite{zwanzig2001nonequilibrium} Here the chosen coarse-grained sites are the $\alpha$-carbon of each amino acid in the protein primary sequence. \textcolor{black}{To obtain a linear Langevin equation,\cite{lamm1986langevin} we take the coordinates tracing the backbone configuration of the protein to be complete of the relevant slow configurational degrees of freedom, and neglect system memory. Inertial terms may be discarded as a protein in aqueous solution is safely in the overdamped limit.} The intramolecular distribution around the folded state is assumed to be Gaussian, and the parameters in the distribution are directly obtained from the starting configurational ensemble.\cite{caballero2007theory,guenza2008theoretical}
The coarse-grained LE4PD represents the balance of viscous dissipation with the entropic restoring force and a random Brownian force due to the random collisions of the coarse-grained protein with the fast-moving projected atoms belonging to solvent, ions, and the protein. The time evolution of the coordinate of the coarse-grained site $i$ is well-described by the following equation 
\begin{eqnarray}
\overline{\zeta}\frac{\partial\vec{R}_i(t)}{\partial t}= - \frac{3 k_B T}{l^2} \sum_{j,k} H_{ij}A_{jk}\vec{R}_k(t)+\vec{F}_i(t) \ ,
\label{LE1}
\end{eqnarray}
 where $k_B$ is the Boltzmann constant, $T$ is the temperature, $l^2$ is the squared bond distance, and $\overline{\zeta}$ is the average monomer friction coefficient, defined as $\overline{\zeta}= N^{-1} \sum_{i=1}^{N} \zeta_i$, with $\zeta_i$ the friction of the monomer $i$. $\vec{F}_i(t)$ is a delta-correlated random force due to projecting the system dynamics onto the coarse-grained sites, where fluctuation-dissipation requires $\langle F_{i\alpha}(t) F_{j\beta}(t')\rangle = 2 k_B T \zeta_i \delta(t-t') \delta_{i,j}\delta_{\alpha,\beta}$ where $\alpha,\beta$ are cartesian indices.
 Eq. \ref{LE1} is the well-known Rouse-Zimm equation for the dynamics of polymers in solution.\cite{bixon1978optimized,doi1986theory}

\textcolor{black}{To obtain an effective linear description we assume a well-folded state where site-site correlations are Gaussian in nature. The structural force matrix $\mathbf{A}$ defines the effective mean-force potential, $V(\{\vec{R}\})=\frac{3 k_B T}{2 l^2}\sum_{i,j=1}^{N}A_{ij}\vec{R}_i\cdot\vec{R}_j$, which has been successfully adopted in theories of protein folding to describe the final state of the folding process.\cite{portman1998variational} The $\mathbf{A}$ matrix is calculated as 
\begin{eqnarray}
\label{matrixA}
\mathbf{A}= \mathbf{M}^T \left( \begin{array}{cc}
0 &\mathbf{0} \\
\mathbf{0} & \mathbf{U} \\ \end{array} \right) \mathbf{M} \ ,
\end{eqnarray}
 where $\mathbf{M}$ is the matrix that defines the center of gyration and the connectivity between sites, $\sum_{j}M_{ij}\vec{R}_j=\vec{l}_i$. In a protein the $\alpha$-carbons are connected linearly, so that for $i > 1$ the matrix is defined as $M_{i,i-1}=-1$ and $M_{i,i}=1$, with $i=2, . . . , N$, while $M_{1,i}=1/N$ for the first row, and $M_{i,j}=0$ otherwise. The $\mathbf{U}$ matrix is the bond correlation matrix with $(\mathbf{U}^{-1})_{ij}=\frac{\langle \vec{l}_i \cdot \vec{l}_j \rangle}{\langle |\vec{l}_i | \rangle \langle | \vec{l}_j | \rangle }$.}

\textcolor{black}{The matrix $\textbf{H}$ is the 
hydrodynamic interaction matrix, which describes the interaction between protein sites occurring through the liquid, represented as a continuum medium. While it is standard to utilize hydrodynamical models to obtain the translational and rotational dynamics of proteins,\cite{garcia2000calculation} the contribution of hydrodynamical effects to protein internal motion is generally neglected. While this may be justified for very localized motion, in general the non-local hydrodynamic coupling alters the timescale and nature of the large-amplitude highly correlated internal motion and cannot be neglected.\cite{granek2011proteins,copperman2014coarse-grained} To maintain an effective linear description, the hydrodynamic interaction must be preaveraged. While the derivation of the hydrodynamic interaction utilizes the Oseen tensor following the general Rouse-Zimm treatment of polymer chains in dilute solution,\cite{doi1986theory} other methods such as the Rotne-Prager interaction tensor reduce to the same form upon preaveraging over the equilibrium ensemble.}\cite{rotne1969variational} The elements in the matrix of the hydrodynamic interaction are defined as 
\begin{eqnarray}
H_{ij}=\frac{\overline{\zeta}}{\zeta_i}\delta_{ij}+(1-\delta_{ij})\overline{r}^w \langle\frac{1}{r_{ij}}  \rangle  \ . 
\end{eqnarray}
where $\overline{r}^w= N^{-1} \sum_{i=1}^{N} r^w_i$ is the average hydrodynamic radius which is defined below. This is a perturbative hydrodynamic interaction accounting for the nature of the amino acid primary structure as a heteropolymer made up of building blocks of different chemical types, propagating through the aqueous solvent but screened in the dense hydrophobic core. The site-specific friction parameters, $\zeta_i$, are obtained by calculating the solvent-exposed surface area, and calculating the total friction of the $i_{th}$ site via a simple extension of Stoke's law as
\begin{eqnarray}
\zeta_i=6\pi(\eta_w r^w_i + \eta_p r^p_i) \ . 
\label{Stoke}
\end{eqnarray}
Here $\eta_w$ and $r^w$ denote, respectively, the viscosity of water and the radius of a spherical bead of identical surface area as the solvent-exposed surface area of the residue, the hydrodynamic radius\cite{caballero2007theory}, while $r^p$ denotes the hydrodynamic radius related to the surface not exposed to the solvent. The internal viscosity is $\eta_p$, which we approximated in our previous work to be related to the water viscosity rescaled by the local energy-barrier scale $\sim k_B T$.\cite{copperman2014coarse-grained,sagnella2000time} The largest possible value of $\overline{r}^w$ that maintains a positive definite solution of the matrix diagonalization is adopted to avoid the well-known issue with the preaveraging of the hydrodynamic interaction in dense systems.\cite{zwanzig1968validity} For example, in the application of the model to HIV protease, the calculated $\overline{r}^w=2.28\AA$ is very close to the adopted value of $\overline{r}^w=2.23\AA$, which avoids negative eigenvalues.

Because we focus only on the bond orientational dynamics and not translation, in the interest of a simpler notation we separate out the zeroth order translational mode from the internal dynamics. Following the same notation introduced for the orientational dynamics of star polymers,\cite{guenza1992reduced} 
we define $\mathbf{a}$ as the $\mathbf{M}$ matrix after suppressing the first row used to define the center of mass, and define $\mathbf{L}=\mathbf{a} \mathbf{H} \mathbf{a}^T$. The orientational  Langevin equation governing the bond dynamics is 
\begin{eqnarray}
\frac{\partial\vec{l}_i (t)}{\partial t}= - \sigma  \sum_{j,k} L_{ij}U_{jk}\vec{l}_k(t)+\vec{v}_i(t) \ ,
\label{LE}
\end{eqnarray}
 with $i,j=1, ..., N-1$, and where $\sigma=3 k_B T/(l^2\overline{\zeta})$, and $\vec{v}_i(t)$ is the random delta-correlated bond velocity. 

\textcolor{black}{Eq. \ref{LE} represents a set of $N-1$ first-order coupled differential equations, which are solved by finding the matrix of eigenvectors $\mathbf{Q}$ which diagonalizes the product of matrices $\mathbf{LU}$. In these diffusive modes we have $N-1$ uncoupled linear equations where each mode is just a sum over the original bond vector basis $\vec{\xi_a}(t)=\sum_{i}Q^{-1}_{ai}\vec{l}_i(t)$.  We define $\lambda_a$ to be the eigenvalues of $\mathbf{LU}$ with $\sum_{i,j,k}Q^{-1}_{ai}L_{ij}U_{jk}Q_{kb}=\delta_{ab}\lambda_a$, ordered from smallest to largest $\lambda$. Like the set of bond vectors $\vec{l}_i(t)$ the set of coordinates $\vec{\xi}_a(t)$ defines the instantaneous conformation of the macromolecule. While the $\mathbf{L}$ and $\mathbf{U}$ matrices are individually symmetric, the $\mathbf{LU}$ matrix is not necessarily symmetric, making the $\mathbf{U}$ matrix only approximately diagonal in the $\mathbf{LU}$ eigenvector basis. The mean squared mode length is then $\langle\xi^2_a\rangle\equiv\frac{l^2}{\mu_a}$ with $\mu_a$ not exactly the eigenvalues of the bond correlation matrix alone, but defined by the sum $\sum_{i,j}Q^{-1}_{ai}U^{-1}_{ij}Q^{-1}_{aj}\equiv\frac{1}{\mu_a}$. The diffusive mode basis spans the same space as the bond vector basis with near linearity: $\langle\vec{\xi}_a \cdot \vec{\xi}_b \rangle\cong\delta_{ab}  l^2/\mu_a$.} 

The first three global modes of the LE4PD describe the rotations of the folded structure as the rotational diffusion tensor. For proteins which have an arbitrary folded structure, the full rotational diffusion equation of an anisotropic 3-dimensional body must be solved. This alters the relaxation of the three global modes describing the rotational relaxation in the inertial lab frame.\cite{favro1960theory,copperman2014coarse-grained} 

To account for the affect of the local energy barriers on the internal dynamics, the friction becomes mode dependent by assuming thermal activation over the mode-dependent energy barrier $\langle E^{\dag}_a \rangle$
\begin{eqnarray}
\overline{\zeta} \rightarrow \overline{\zeta}  \ exp[\langle E_a^{\dag} \rangle/(k_B T)] \ ,
\end{eqnarray}
leading to the slowing of the mode timescale $\tau_a = \frac{l^2 \overline{\zeta}}{3 k_B T \lambda_a }$ by
$\tau_a \rightarrow \tau_a  \ exp[\langle E_a^{\dag} \rangle/(k_B T)]$. \textcolor{black}{The energy barriers in the modes $\langle E^{\dag}_a \rangle$ can be calculated directly from an MD simulation trajectory when available, and are found to be related to the mode cooperativity, as discussed in section \ref{FEL}.}

 This simple dynamical renormalization provides an average correction to the dynamics of the Langevin Equation, which approximately accounts for the local barrier crossing, and is in agreement with free energy landscape theories suggesting activated dynamics.\cite{bryngelson1995funnels,onuchic1997theory}
 As a first approximation, the depth of the minimum free-energy well in the mode serves as the relevant barrier to transport.

\subsection{Local Dynamics}
To provide a further assessment of the accuracy of the LE4PD for the local dynamics we compare its theoretical predictions with experimental data of NMR $T_1$, $T_2$, and NOE relaxation.  The  physical quantities of interest for this test are the bond autocorrelation function and the second order Legendre polynomial of the time dependent bond orientation.

For each bond $i$ along the backbone of the protein, the bond autocorrelation function is defined in the formalism of the Langevin equation as 
\begin{eqnarray}
M_{1,i}(t)=\frac{\langle\vec{l}_i(t)\cdot\vec{l}_i(0)\rangle}{\langle l_i^2 \rangle}= & \sum_{a=1}^{N-1} A_{ia}\exp[- t/\tau_a] \ ,
\end{eqnarray}
with $A_{ia}=\frac{Q_{ia}^2}{\mu_a}$ and $\tau_a$ the correlation time for the $a$th mode.

Another quantity of interest is the second order Legendre polynomial of the time dependent bond orientation function $P_2(t)=\frac{3}{2}\langle\cos^2[\theta(t)]\rangle-\frac{1}{2}$ which can be related to the first order bond autocorrelation by
\begin{eqnarray}
P_{2,i}(t)=1-3(x^2-\frac{2}{\pi}x^3(1-\frac{2}{\pi}\arctan x)) \ ,
\label{p2theory}
\end{eqnarray}
which is a function of $M_{1,i}(t)$ as 
$x=\frac{[1-M_{1,i}(t)^2]^{\frac{1}{2}}}{M_{1,i}(t)}$.

This expression relies on assuming a Gaussian form for the joint probabilities in normal mode coordinates.\cite{perico1985viscoelastic} 
For dipolar relaxation, the Fourier transform of $P_{2,i}(t)$ defines the spectral density 
from which spin-lattice ($T_1$) and spin-spin ($T_2$) relaxation times, and nuclear Overhauser effect (NOE) can be calculated and compared with NMR measurements.\cite{copperman2014coarse-grained}

\section{Structural ensembles of proteins}
The LE4PD model predicts the dynamics of the protein using the structural ensemble as input. By generating a structural ensemble through relatively short time ($\sim10 \ ns$) MD simulations, the needed input for the LE4PD was evaluated leading to accurate predictions for the global and site-specific dynamics of Ubiquitin\cite{copperman2014coarse-grained} and the signal transduction protein CheY.\cite{caballero2007theory} The accuracy of the simulations, however, depends upon the accuracy of the force-field used, and sampling the full configurational space 
\textcolor{black}{can become computationally expensive depending upon the size of the protein and the extent of configurational rearrangements.}

\subsection{\textcolor{black}{Building statistical ensembles from metastable configurations}}
\textcolor{black}{We take as an ansatz that the configurational space of a folded protein is spanned by limited number of conformational states, and that these conformational states are known \textit{a priori}. As an alternative procedure to performing MD simulations we assume as starting configurational ensembles the conformers that were measured experimentally by NMR. The extent to which NMR solution structure conformers represent important metastable states of the protein, as opposed to uncertainty due to incomplete experimental information, is controversial and varies between different NMR structures.\cite{spronk2003precision} It is certainly clear that NMR structural ensembles do encode some measure of the conformational variability of the protein, as NMR structural ensembles have been shown to correlate highly with structural ensembles generated by MD simulation,\cite{jamroz2014cabs} and have been used to gain valuable insight into protein flexibility in computational studies of ligand binding.\cite{damm2007exploring} In our model we investigate the assumption that all conformers represent metastable protein configurations which contribute equally to the full ensemble, and use the resulting dynamical predictions to evaluate the ability of the input structural ensemble to span the experimentally observed dynamics.} 

Fluctuations around the local conformational states are imposed by applying a Gaussian Network Model (GNM).\cite{bahar1997direct} \textcolor{black}{While many elastic network models of varying complexity are in routine use, the differences in the predicted local flexibilities are usually small and affect only the short-time dynamics in the picosecond regime.} The GNM builds a harmonic network of interactions around each residue based on a distance cutoff criteria, and solves the resulting site-site fluctuations as a linear matrix equation. GNM models have been shown to reproduce well crystalline state fluctuations measured as Debye-Waller Temperature factors (B-factors) and thus are a good representation of the short-time fluctuations while they require minimal computational effort. Once combined with the 
LE4PD the theory provides a realistic and computationally inexpensive prediction of the dynamics of proteins on a wide range of time scales, from the local fluctuations to the large, concerted, conformational transitions. 

From the GNM we define the bond correlation matrix locally around the $\alpha$th conformer $U_{\alpha,ij}$. The GNM defines the pairwise fluctuations 
$\langle\vec{\Delta R}_i \cdot \vec{\Delta R}_j \rangle=\frac{3 k_B T}{\gamma}\Gamma^{-1}_{ij}$
where $\mathbf\Gamma$ is the Kirchoff adjacency matrix defined using a cutoff radius of $7.0\AA$\cite{bahar1997direct,atilgan2001anisotropy} and $\gamma$ is the harmonic interaction strength. We found that in general a value of $\gamma=0.06\frac{kcal}{mol \AA^2}$ is needed to match the short-time $1 \ - \ 10 \ ps$ orientational fluctuations of the protein from the MD simulation.

 An interaction strength of $\sim1\frac{kcal}{mol \AA^2}$ is typically used with the GNM to predict crystallographic B-factors; this order of magnitude difference in interaction strength may be due to the local anharmonic softening of the orientational potential energy surface due to the aqueous solvent.\cite{levy2006water} The boundary water layer of hydrated proteins in aqueous solution is highly mobile in the picosecond regime\cite{pal2002biological}; the constant shifting of the protein-water hydrogen bonds may lead to enhanced orientational fluctuations which are completely local in nature. This effect is absent in the crystalline state, where the hydration water is much more static. 

Recognizing that in the body-fixed reference frame $\vec{l}_i(t)-\langle\vec{l}_i\rangle=[\vec{R}_{i+1}(t)-\vec{R}_{i}(t)]-[\langle\vec{R}_{i+1}\rangle-\langle\vec{R}_{i}\rangle]$ we can determine the local bond correlation matrix around each conformer as
\begin{eqnarray}
(\mathbf{U})^{-1}_{\alpha,ij} & = & \frac{1}{\langle |\vec{l}_i | \rangle \langle | \vec{l}_j | \rangle }\bigg[\langle\vec{l}_i\rangle\cdot\langle\vec{l}_j\rangle+  \frac{3 k_B T}{\gamma} \nonumber 
 \\
&& (\Gamma^{-1}_{ij}  +  \Gamma^{-1}_{i+1,j+1}-\Gamma^{-1}_{i,j+1}-\Gamma^{-1}_{i+1,j})\bigg] \ .
\end{eqnarray}
The total $\mathbf{U}$ matrix is then simply the average $U_{jk}=\frac{1}{N_c}\sum_{\alpha=1}^{N_c}U_{\alpha,jk}$ with $N_c$ the number of conformers in the NMR structural ensemble. Similarly, the hydrodynamic matrix $\mathbf{H}$, the site friction coefficient $\zeta_i$, and all other input quantities to the LE4PD, are calculated separately for each conformer and then the statistical average is taken over all the conformers. 
This is an extremely simplistic picture of the structural ensemble of a protein; however the dynamics predicted by this structural ensemble generated by the set of NMR conformers is consistent in many ways with the much more detailed ensemble generated through the sophisticated process of explicit solvent MD simulation. The set of NMR conformers provides us an ensemble of important metastable structural minima in the free energy landscape; and the GNM provides fluctuations around these minima. Molecular anisotropy, rotational diffusion, hydrodynamic interactions, and local energy barriers are included through the LE4PD.

\subsection{Building statistical ensembles from Molecular Dynamics simulations}
Because both the determination of NMR conformers and the experimental measurements of NMR relaxation are affected by errors, we performed as a further test MD simulations of the same systems to evaluate the quality of agreement of the LE4PD starting from the NMR and the MD conformers. Simulations were performed in explicit solvent using the spc/e water model. We utilized the AMBER99SB-ILDN\cite{lindorff2010improved} atomic force field for proteins and the GROMACS\cite{gromacs1,gromacs2,gromacs3,gromacs4} molecular dynamics engine was utilized on the TRESTLES supercomputer at San Diego.\cite{towns2014xsede} All system conditions, e.g. temperature and salt concentration, were set to reproduce the experimental conditions. The systems were solvated and energy minimized, and then underwent a $500 \ ps$ tempering and equilibration routine including pressure coupling. The production simulations were performed in the canonical ensemble, using a velocity rescaling thermostat.\cite{bussi2007canonical}

For the PR95 protease monomer, simulations were performed starting from each of the twenty conformers in the NMR structure, resulting in a set of twenty production ensembles used as input to the LE4PD, with averages taken over all twenty results. The same set of production trajectories for Ubiquitin were used from our previous work.\cite{copperman2014coarse-grained} For the remaining five proteins, the first conformer was chosen as the starting structure, and only one simulation was performed for each protein. Each simulation had $50 \ ns$ of production. For each trajectory the root mean square deviation (RMSD) was calculated and statistics were only collected in the equilibrated sections of the trajectory. The trajectories were also required to contain only reversible transitions, as monitored by the RMSD. The simulation time effectively used in the LE4PD for each trajectory ranged from $10$ to $30 \ ns$. The simulations performed and the protein databank structures used are summarized in Table \ref{proteintable}.

\begin{minipage}{\linewidth}
\centering
 \captionof{table}{Systems and Structural Ensembles}\label{proteintable}
   \begin{tabular}{ccccc} \hline \hline
Protein & MD Sim.  & Starting Struct. & Temp. & NMR + GNM\\
\hline
Protease	 	& 20 x 50 ns	& 1Q9P (1-20)	& 293K	& 1Q9P (1-20)	\\
1GF2R 			& 160 ns	& 2M6T (1)		& 273K	& 2M6T (1-20)	\\
N-TIMP-1 		& 50 ns		& 1D2B (1)		& 293K	& 1D2B (1-30)	\\
S836 			& 50 ns		& 2JUA (1)		& 298K	& 2JUA (1-20)	\\
CPB1 			& 50 ns		& 1MX7 (1)		& 298K	& 1MX7 (1-22)	\\
KAPP 			& 50 ns		& 1MZK (1)		& 298K	& 1MZK (1-30)	\\
Ubiquitin		& 10 x 10ns	& 1UBQ (1)		& 300K	& 1XQQ (1-128)	\\
\hline
    \end{tabular}
\end{minipage}

\begin{figure}[htb] 
\includegraphics[width=.8\columnwidth]{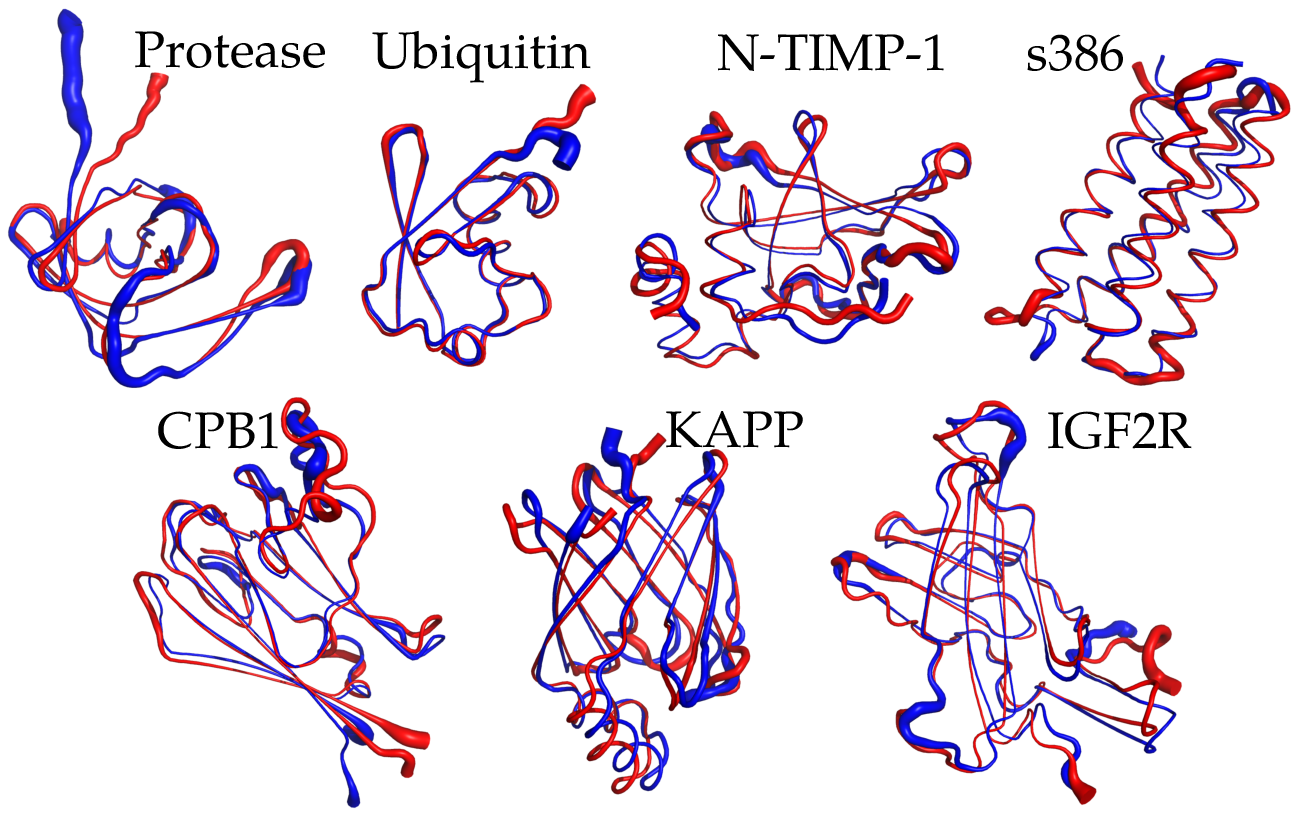}
\caption{Average configuration from the MD simulation ensemble (red) and from the NMR structural ensemble (blue), with the thickness of the ribbon accurate to the local orientational distribution.}
\label{startingconf}
\end{figure}

The configurational ensembles that emerge from the NMR ensembles and from the MD simulations are reported in Figure \ref{startingconf}. For all but the Ubiquitin protein, the starting configuration was from the NMR structure, yet the \textcolor{black}{equilibrated simulation conformations do not exactly resemble this initial structure.} Overall, however, the global fold is fully preserved, and the conformational differences are specific in nature. This indicates the NMR structures were not necessarily in an exact free energy minimum of the AMBER force-field model, though this does not indicate whether the MD equilibrated protein structures are necessarily more accurate. For all but the s836 protein, the structural variation in the NMR ensembles is slightly larger than the MD simulation. This is primarily true in the intrinsically disordered regions of the protein, such as the C-terminal and N-terminal tails. This 
\textcolor{black}{may be} because the limited simulation times do not fully sample the configurational space. \textcolor{black}{A study over a test set of 140 proteins found high correlation between the fluctuations of NMR ensembles and MD simulations, and found that the increased sampling allowed by using a coarse-grained protein model led to even higher correlation between simulations and NMR ensembles.\cite{jamroz2014cabs}  What does agree quite remarkably are the locations of enhanced flexibility and the timescales of the motion, which can} be seen in Figures \ref{nmrfromMD} and \ref{nmrfromNMR}, showing the calculated NMR relaxation times from the ensembles.

\begin{figure}[htb] 
\includegraphics[width=.8\columnwidth]{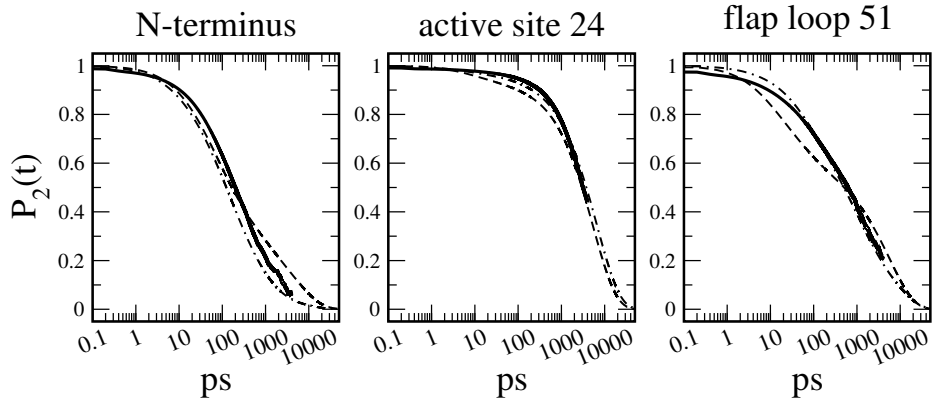}
\caption{$P_{2,i}(t)$ time correlation function 3 different sites along the protein sequence of the HIV protease monomer PR95, calculated directly from the conformer simulations (solid line), from the LE4PD theory with the conformer simulations as input (dashed line), and from the LE4PD with the NMR conformer ensemble as input (dashed-dotted line).}
\label{p2}
\end{figure}

To evaluate the consistency between the dynamics generated using the MD ensembles as input, and the NMR conformer ensembles, we compare the full decay of the $P_{2,i}$ correlation function of the $i$th $C_\alpha$-$C_\alpha$ segment in the HIV protease protein, with the data from simulations (see Figure \ref{p2}). While there are many differences between the analytical predictions from the NMR ensemble and the MD ensemble, and differences especially at short times, overall Figure \ref{p2} shows that the agreement is quite good as the LE4PD, from both ensembles, can model quite accurately the site-specific internal and rotational dynamics of the protein.

\section{Dynamical barriers and cooperativity}
\label{FEL}
Analysis of the collective fluctuations obtained from simulations of proteins \cite{amadei1999kinetic} has shown that the dynamics around the minima of energy is well described by small fluctuations inside metastable states at low local energy and by the crossings between them. By reverting the LE4PD equation to its mode form, the structural representations of these important metastable minima can be identified as a function of mode number. We investigate the nature of the free energy surface of a protein around its folded ground state.

Each diffusive mode obtained from the diagonalization of Eq. \ref{LE} is a vector defined by the linear combination of the bond vectors weighted by the eigenvectors of the product of matrices $\mathbf{LU}$, as $\vec{\xi}_a(t)=\sum_{i}Q^{-1}_{ai}\vec{l}_i(t)$. In polar coordinates the vector is represented as $\vec{\xi}_a(t)= \{| \vec{\xi}_a(t)|, \theta_a(t), \phi_a(t) \}$. The most relevant changes in the diffusive mode free energy occur as the angles, expressed in the spherical coordinates, span the configurational space. For any diffusive mode $a$, the free energy surface is defined as a function of the spherical coordinate angles $\theta_a$ and $\phi_a$ as  $F(\theta_a,\phi_a)= -k_B T \log \left \{P(\theta_a,\phi_a) \right\}$,with $P(\theta_a,\phi_a)$ the probability of finding the diffusive mode vector having the given value of the solid angle. Given that we are interested in the explicit representation of the structure at the minima of interest, all structures from the simulation ensemble which pertain to a particular $\theta,\phi$ orientation, which is a relatively deep minima in the mode free energy, are extracted and averaged.

\begin{figure}[htb] 
\includegraphics[width=.9\columnwidth]{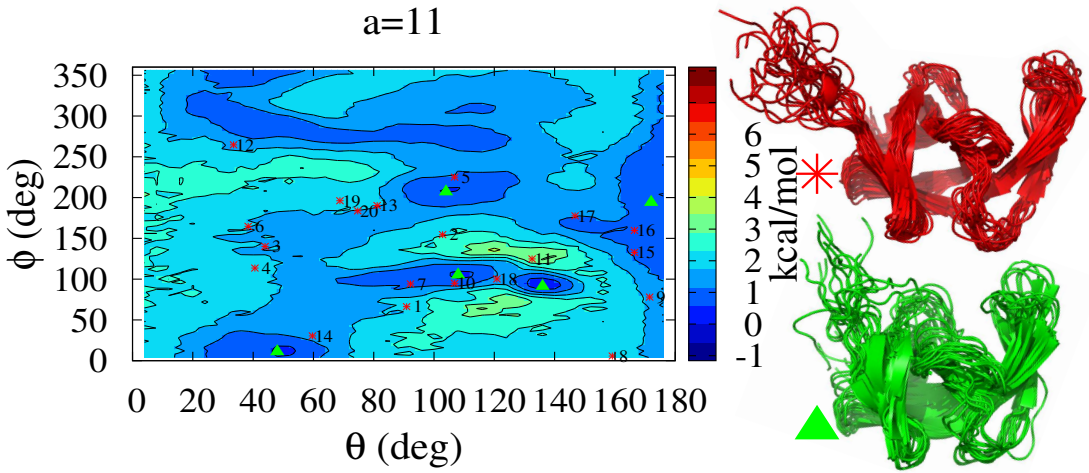}
\caption{Internal mode $a=11$ free energy surface of the HIV protease ($a=4$ is the first internal mode) on the left. Projections of the NMR conformer structures, labeled by conformer index, are plotted with red stars, and the simulation minima from which structures were calculated are marked on the free energy surface with green triangles. Structural minima from simulation modes $a=4-20$ are on the bottom right, and the set of NMR conformers on the top right.}
\label{modes}
\end{figure}

By calculating the average structure at each minima we obtain the structural ensemble of metastable states spanning each internal mode of fluctuation for the protein. As a representative example, the free energy landscape in the LE4PD modes from the MD simulation of the HIV protease monomeric construct is presented in Figure \ref{modes}. The ensemble of structural minima on the mode free energy surfaces generated from MD simulations, and the structural ensembles directly measured by NMR experiments, are compared as well. The full configurational landscape for each mode is generated from the combination of twenty well-equilibrated independent simulation trajectories. Each trajectory starts from a different experimental NMR conformer and runs to $50 \ ns$ of simulation time. Superimposed to the full configurational landscape from simulations, the twenty starting configurations measured experimentally by NMR are reported as red stars. 
\textcolor{black}{The combination of the trajectories creates a complex free energy landscape, which is only partially spanned by the NMR conformers.} The starting NMR configurations are often close to energy minima (reported as green triangles), but they do not exactly correspond to them. Nor they are fully representative of all the minima that define the configurational landscape obtained from the simulation trajectories.

\begin{figure}[htb] 
\includegraphics[width=.8\columnwidth]{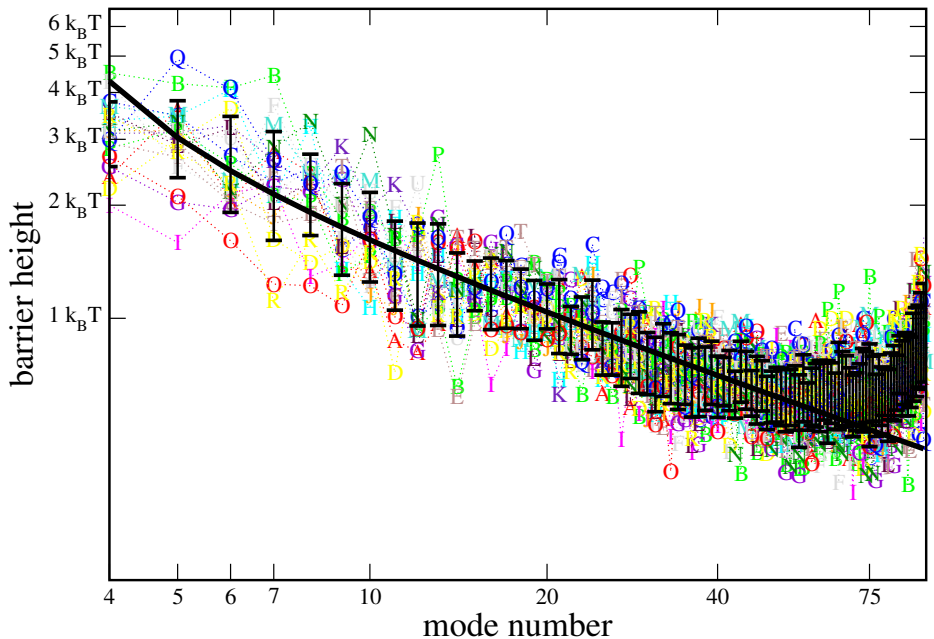}
\caption{The free energy barrier to diffusion in the modes $E^\dag_a$ calculated directly from the set of 20 simulations of the HIV protease at $T=293K$, indexed by letter A-T, with standard deviations around the average as black bars. The black line is the fit to the scaling form $E^\dag_a \propto (a-3)^{-.50}$.}
\label{barriers}
\end{figure}

The fluctuations in each mode appear to be spanned by a handful of metastable minima. As the mode index increases the fluctuations progress from collective in nature to more local. 
The typical energy barrier in each mode, $a$, is evaluated from the simulation as the Median Absolute Deviation\cite{ruppert2010statistics} from the global minimum $E_{gs}$, that is $E^{\dag}_a=median_{(\theta,\phi)}(E_a(\theta,\phi)-E_{gs,a})$. The depth of these minima, or the barriers between them, are largest for the low mode numbers corresponding to the most collective, large-amplitude fluctuations. 
 Figure \ref{barriers} shows that the energy barriers $E^\dag_a$ as observed in the simulation trajectory can be well described as scaling with the mode index, a measure of the mode cooperativity, over a large range of the protein fluctuations.  The observer scaling with mode number follows $E^\dag_a\propto (a-3)^{-.5}$, where  the first three rotational modes have been separated out. At a local enough length scale, where the specific chemical nature of the amino acid is most important, the energy barriers are no longer described by this expression. 

The observed scaling law is consistent with the hierarchical nature of the protein free energy landscape.  Each mode describes dynamics involving a number of bonds in the protein, which need to move collectively in a cooperative fashion. At short times the bonds fluctuate independently, while  large-amplitude correlated fluctuations occur when all the bonds transition collectively.\cite{jackson1993time}
The equilibrium probability for the $Z$ gating bonds to independently transition away from the "correct" orientation with energy preference $E$ is $P(Z) \propto \exp(-\frac{ZE}{k_B T})$. In a transition state perspective this can be interpreted as a free energy barrier which scales proportionally with the number of bonds cooperatively rearranging. This model is similar to the Adam-Gibbs theory of the glass transition,\cite{adam1965temperature,starr2013relationship} relating
the complex hierarchical nature of the free energy landscape the protein in solution to a structured glassy fluid.\cite{onuchic1997theory,wales2015perspective} 
The observed scaling form is included in the simulation-free LE4PD approach, which adopts the set of NMR conformers as the input structural ensemble.

\section{Predictions of NMR relaxation are compared to experiments}
Theoretical predictions for $P_{2,i}(t)=\frac{1}{2}(3\cos^2{\theta_i(t)}-1)$ are obtained from $M_{1,i}(t)$ using Eq. \ref{p2theory}, and are used to calculate $T_1$ and $T_2$ relaxation times, and NOE, which are measured experimentally. 15N NMR backbone relaxation experiments are very sensitive to the site-specific dynamics in the picosecond to the nanosecond regimes.\cite{palmer2001nmr}
To test the LE4PD approach using the NMR solution structures to generate the structural ensemble, we constructed dynamical models for seven proteins for which NMR relaxation data and NMR solution structures were available. These proteins were N-TIMP-1 (1D2B)\cite{gao2000tissue}, a de Novo $\alpha$-helix bundle protein s836 (2JUA)\cite{go2008structure}, Cellular retinol-binding protein I CPB1 (1MX7)\cite{lu2003two},  Kinase-associated protein phosphatase KAPP (1MZK)\cite{lee2003nmr,lee20031H}, Insulin Growth Factor 2 Receptor IFG2R domain 11 (2M6T)\cite{williams2007structural}, Ubiquitin (1UBQ)\cite{vijay1987structure,lienin1998anisotropic}, and HIV Protease monomer (1Q9P).\cite{ishima2001folded,ishima2003solution} 

\begin{figure}[htb] 
\includegraphics[width=.9\columnwidth]{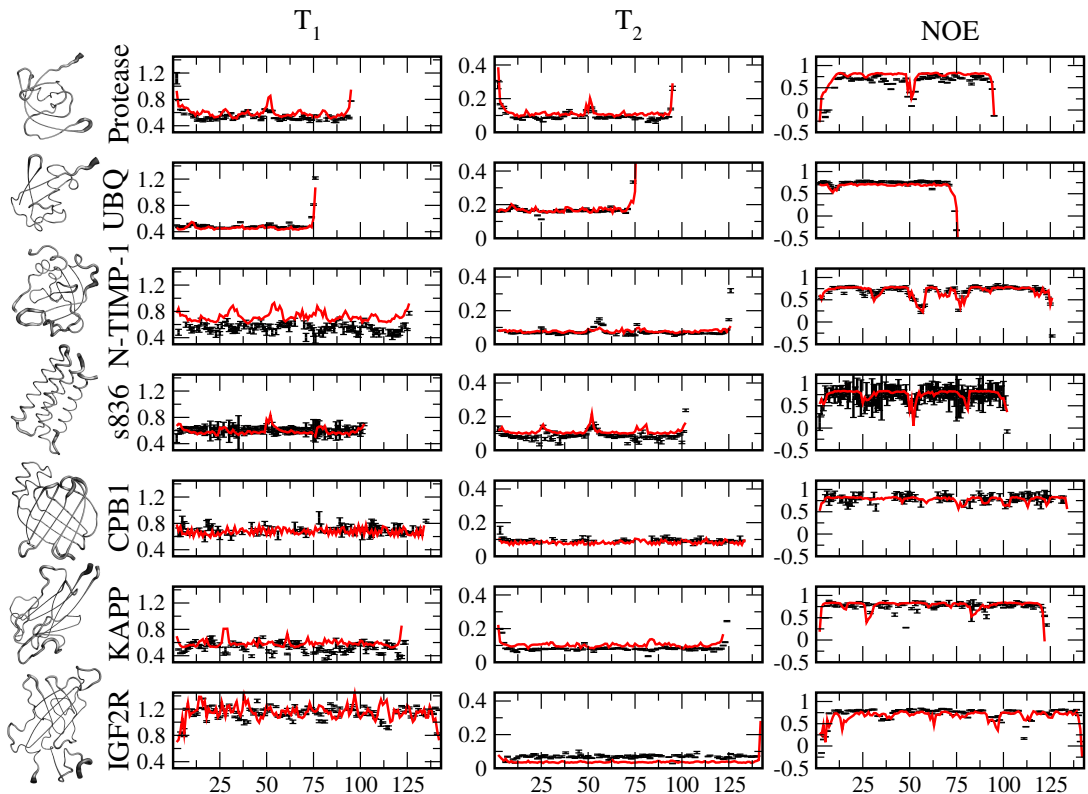}
\caption{$T_1$, $T_2$, and $NOE$ relaxation times (see Table 1) for seven different proteins. Comparison between experimental (black) and theoretical values from LE4PD theory from MD generated ensembles (red).}
\label{nmrfromMD}
\end{figure}

\begin{figure}[htb] 
\includegraphics[width=.9\columnwidth]{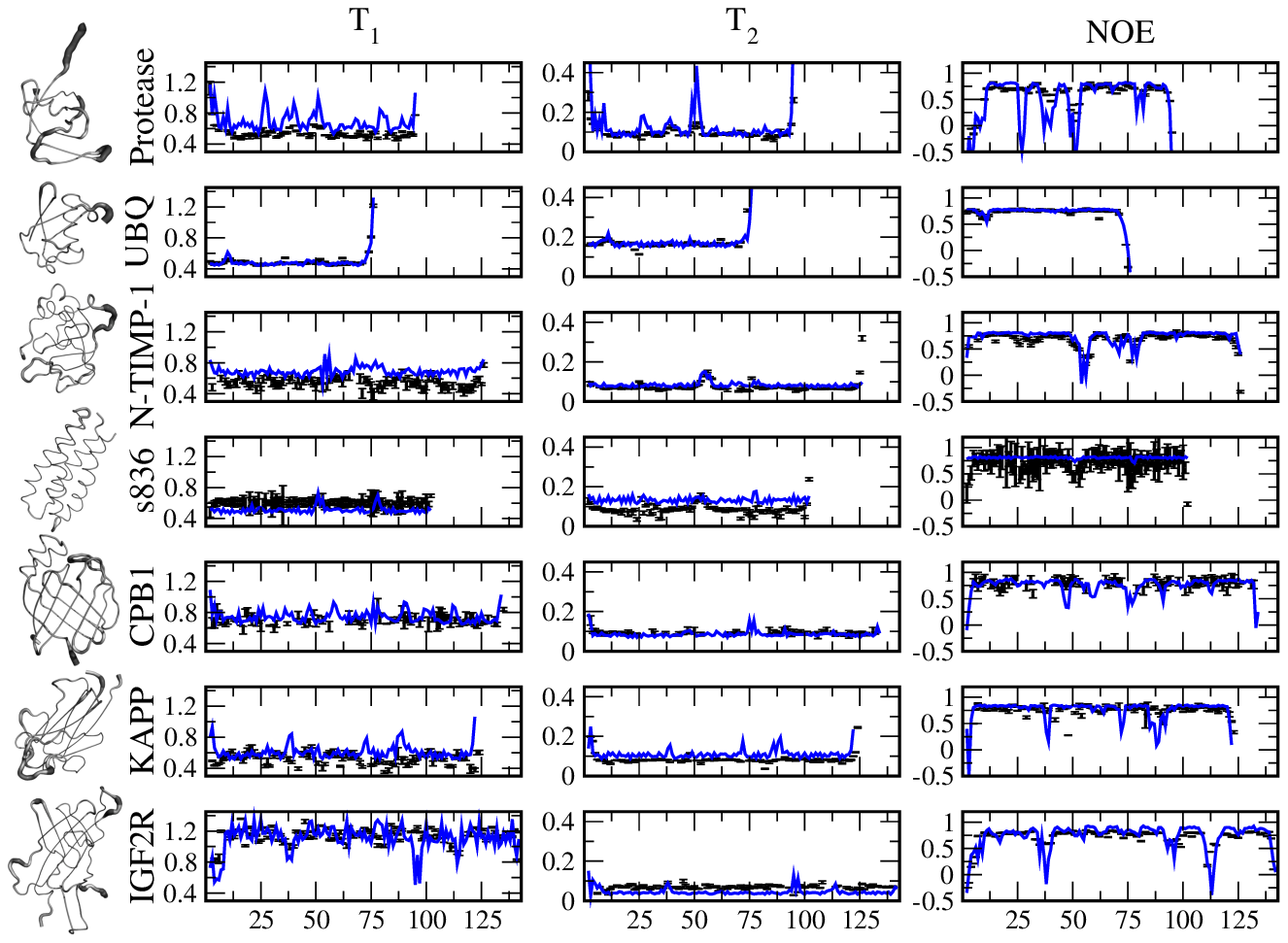}
\caption{$T_1$, $T_2$, and $NOE$ relaxation times (see Table 1) for seven different proteins. Comparison between experimental (black) and theoretical values from LE4PD theory from ensembles generated from NMR conformers (blue).}
\label{nmrfromNMR}
\end{figure}

The input parameters to the LE4PD equation change from protein to protein: the structural parameters such as bond length, monomer friction, hydrodynamic radius, and the pairwise bond correlations are determined from the structural ensemble, while the thermodynamic parameters such as solvent viscosity, and temperature, are defined by the experimental conditions. 
The viscosity was set to account for temperature dependence and content of deuterated water.\cite{wong2008evaluating} Parameters such as the protein internal viscosity $\eta_p$, the 
proportionality constant between cooperativity and energy barriers, and the characteristic parameters needed to calculate the NMR relaxation times, such as the chemical shift or $\langle1/r_{NH}^3\rangle$, were assumed to be identical for all proteins in this study and identical to those used in our previous work.\cite{copperman2014coarse-grained}

Figure \ref{nmrfromMD} and \ref{nmrfromNMR} displays the calculations of $T_1$, $T_2$ and NOE relaxation times as they are directly predicted by the LE4PD approach and the NMR experimental data. NMR experimental data of relaxation times are not used at all in any point to optimize the theoretical calculations, so these are independent theoretical predictions. The comparison between theory and experiments is performed for each amino acid in the protein and reported as a function of the protein primary sequence. Also reported are the experimental uncertainties for the NMR data of each protein.

\begin{minipage}{\linewidth}
\centering
 \captionof{table}{Correlation of the LE4PD with Experimental Data of NMR Relaxation}\label{protnmrtable}
   \begin{tabular}{cccccc} \hline \hline
    Protein & $\rho_{total}$ & $\rho_{T1}$ & $\rho_{T2}$ & $\rho_{NOE}$ & Rel. Err. \\
\hline
    Combined (MD)	& .95	& .88	& .73	& .70	& 20.8\% \\
    Combined (NMR) 	& .93	& .83	& .58	& .69	& 24.9\% \\
    HIV Protease (MD)	& .92	& .73	& .91	& .91	& 32.9\% \\
    HIV Protease (NMR)	& .83	& .65	& .90	& .77	& 42.6\% \\
    Ubiquitin (MD) 	& .98	& .96	& .94	& .97	& 7.0\% \\
    Ubiquitin (NMR) 	& .97	& .96	& .94	& .99	& 7.2\% \\
    N-TIMP-1 (MD) 	& .92	& -.10	& .50	& .82	& 20.1\% \\
    N-TIMP-1 (NMR) 	& .96	& -.18	& .57	& .62	& 22.2\% \\
    s836 (MD) 		& .97	& .03	& .48	& .57	& 20.6\% \\
    s836 (NMR)		& .93	& .18	& .13	& .33	& 34.2\% \\
    KAPP (MD) 		& .96	& .02	& .60	& .60	& 20.1\% \\
    KAPP (NMR) 		& .91	& -.12	& .59	& .46	& 24.6\% \\
    CPB1 (MD) 		& .98	& .03	& .06	& .16	& 9.5\% \\
    CPB1 (NMR) 		& .96	& .03	& .14	& .28	& 10.0\% \\
    IGF2R (MD) 		& .97	& -.06	& .84	& .80	& 24.6\% \\
    IGF2R (NMR) 	& .95	& .34	& .15	& .67	& 25.7\% \\
\hline
    \end{tabular}
\end{minipage}

\begin{figure}[htb] 
\includegraphics[width=.8\columnwidth]{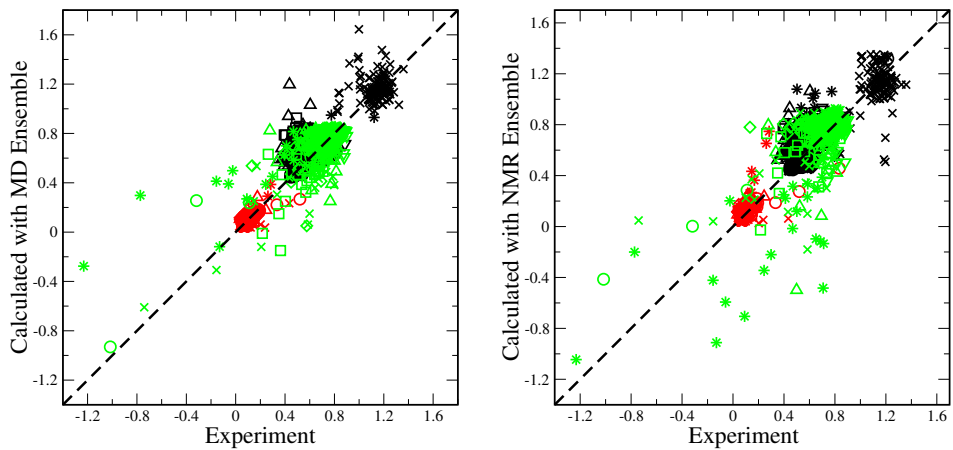}
\caption{Correlation between experimental and calculated values from MD ensembles and NMR ensembles for a set of 7 different proteins. $T_1$ measurements in black, $T_2$ measurements in red, and $NOE$ measurements in green, for Ubiquitin (circles), N-TIMP-1 (squares), s386 (diamonds), CPB1 (downward triangles), KAPP (upward triangles), IGFR2 (x), and HIV protease monomer (star).}
\label{scatter}
\end{figure}

The correlation and errors of the model, using both the MD and NMR solution structures as ensembles, are shown in Table 2. Over this set of $1876$ measurements the overall correlation to the experimental values was similar for both the dynamical models constructed from NMR conformer ensembles or the MD ensembles, but with 17\% lower relative error for the MD derived ensembles than the NMR conformer ensembles. Figures \ref{nmrfromMD} and \ref{nmrfromNMR}, and Table 2, in general show that MD simulations have the most detailed agreement along the primary sequence. The correlation to NOE and $T_1$ are similar, but higher $T_2$ correlation for the MD ensembles.
 Over the seven proteins, the quality of the experimental measurements varies greatly; for example, in the measured relaxation for the s836 protein the experimental values themselves come with $\sim30\%$ error, so that the low correlation of the theory with the experimental data is expected. For the CPB1 protein the experimental measurements in most loop and termini regions were unavailable; this is where the largest variability in the dynamics occurs and where it is possible to develop strong correlation.
 In general, the NOE measurements display the largest site-specific variability along the protein sequence and the highest correlation between theory and experiment for each individual protein. A scatter plot of the calculated and experimental data is shown in Figure \ref{scatter}. The agreement between theoretical predictions and measured NMR is supporting the quality of the predictions of the LE4PD approach.

\textcolor{black}{Because the accuracy of a given NMR solution structure ensemble to represent the conformational diversity of the protein is unknown, the dynamical model built using the LE4PD approach may be useful to evaluate the quality of an available structural ensemble. We apply the method to 9 different NMR structural ensembles of the Ubiquitin protein, PDB codes 1XQQ,\cite{lindorff2005simultaneous} 2KOX,\cite{fenwick2011weak} 2LJ5,\cite{montalvao2012determination} 2NR2,\cite{richter2007mumo} 1D3Z,\cite{cornilescu1998validation} 1G6J,\cite{babu2001validation} 2KLG,\cite{madl2009use} 2MJB,\cite{maltsev2014improved} and 2K39.\cite{lange2008recognition} The comparison to the calculated NMR backbone relaxation in table \ref{ubqnmrtable} shows that all the ensembles capture the primary $T_1$, $T_2$, and $NOE$ baselines and the enhanced flexibility of the tail region. Ensembles 1XQQ and 2NR2 have the highest correlation and lowest relative error; when the unstructured C-tail is not considered in the calculation of the correlation coefficients, it can be seen that the 1XQQ, 2NR2, 2KOX, 2LJ5 and 2K39 ensemble separate as capturing the structural variability of both the C-tail and the more structured portion of the protein, see column 2 of table \ref{ubqnmrtable}. The primary contribution to this correlation comes from the structural variability at the loop containing lysine 6 and 11, important poly-ubiquitination linkage sites involved in cell-cycle control and DNA repair.\cite{komander2009emerging} The structural ensemble generated by molecular dynamics simulation starting from the 1UBQ\cite{vijay1987structure} crystal structure, with results reported in our previous paper,\cite{copperman2014coarse-grained} is perhaps slightly more accurate overall, but only by a very small amount due to considering the correlation without contributions from the C-tail}.

\textcolor{black}{In generating the 1XQQ ensemble the NMR-derived $S_2$ order parameters from the model-free analysis of Lipari and Szabo\cite{lipari1982model} were used as an additional set of restraints in the generation of the ensemble. It is not surprising then that this leads to an accurate dynamical model. We have shown previously that the site-specific variability in model-free derived $S_2$ order parameters correlates strongly with our results,\cite{caballero2007theory} despite differences in the nature of the predicted internal dynamics. This illustrates the complementary utility of the LE4PD approach, which provides a highly detailed model and additional insight beyond that available when performing only a model-free analysis of NMR backbone relaxation.}

\textcolor{black}{The Ubiquitin ensemble 2K39 was constructed to represent the protein fluctuations in only the long-time regime beyond the global correlation time. As such this ensemble is not as accurate overall, and the dynamical model leads to high error and in particular a poor representation of the C-tail dynamics. However, we do see a separation in the mode timescales, with a slow internal process emerging on the order of $\sim400ns$ and with $\rho_{NOE,(res 1-71)}=.71$, suggesting that this ensemble has captured fluctuations in the difficult to access time regime between the global correlation time and the millisecond time regime of conformational exchange. The authors showed that this ensemble spanned the set of known bound Ubiquitin conformations, suggesting that there are configurational fluctuations of the Ubiquitin protein in the many nanosecond regime beyond the global correlation time which are relevant for the recognition of binding partners.\cite{lange2008recognition}}

\begin{table}
 \caption{Correlation of the LE4PD with Experimental Data of NMR Relaxation for Ubiquitin}
   \begin{tabular}{cccccc} \hline \hline
    NMR Conformer & $\rho_{NOE}$ (1-71) & $\rho_{NOE}$ (all res.) & $\rho_{T2}$ & $\rho_{T1}$ & Rel. Error \\
\hline
    MD Sim.		& .71	& .96	& .94 	& .97	& 7.0\% \\
    1XQQ		& .66	& .99	& .94 	& .96	& 7.2\% \\
    2NR2		& .52	& .98	& .94	& .95	& 7.3\% \\
    2LJ5		& .56	& .93	& -.33	& .97	& 7.7\% \\
    2K0X		& .61	& .94	& .80 	& .88	& 8.2\% \\
    1D3Z		& .02	& .88	& .80 	& .94	& 8.2\% \\
    2K39		& .70	& .88	& -.63 	& .93	& 11.0\% \\
    2MJB		& -.01	& .96	& .96	& .92	& 11.2\% \\
    1G6J		& .02	& .92	& .94 	& .92	& 11.5\% \\
    2KLG		& -.05	& .86	& .92 	& .73	& 14.9\% \\

\hline
    \end{tabular}
\label{ubqnmrtable}
\end{table}

\section{Conclusions}
The LE4PD approach was tested across a set of seven different proteins with overall consistent results for both the MD generated ensembles and the NMR conformer ensembles, with an overall correlation to the $1876$ relaxation measurements of $\rho>.93$. \textcolor{black}{Calculations using 9 different available NMR structural ensembles for the ubiquitin protein show that results are strongly dependent upon the quality of the input structural ensemble and experimental data}, and suggest that this approach may be used as a tool to evaluate the quality of a structural ensemble to represent the important protein fluctuations around the ground folded state. 

The consistent results between the MD generated ensembles and the NMR ensembles suggest that protein configurational space around the folded state can be defined by a small set of important metastable minima. However, when determining the dynamics of transitions between these minima, the hierarchical nature of the protein free energy landscape needs to be taken into account. The mode approach of the LE4PD allows one to conveniently separate contributions to the dynamics depending on the timescales involved. The LE4PD prediction of the existence of a barrier height distribution for the dynamics of folded proteins 
is consistent with the physics of glass-forming systems. 

\textcolor{black}{Building a dynamical model from NMR conformer structures using the LE4PD requires only a few seconds to a few minutes on a single processor with a standard desktop computer, with the computational time depending on the size of the protein and on the number of conformers in the NMR solution structure. While explicit solvent atomistic classical MD simulations are well-developed and can be quite accurate, achieving MD simulations with converged dynamics on the same timescale would require on the order of $10,000-100,000$ hours of processor time or more. The LE4PD is not a replacement for MD simulation as a computational method to predict the fluctuations and dynamics of proteins, but it is a useful tool to quickly provide a prediction of the dynamics given an input structural ensemble.}

Even though the simulation-free LE4PD requires minimal computation it is site-specific, informed of intramolecular energy barriers, hydrodynamics, and long-range correlated motion. It is a sophisticated model of protein dynamics and because of its accuracy in predicting the dynamics, with no input from the dynamical data, LE4PD is a valuable and computationally convenient model to investigate barrier-crossing processes on the suite of timescales defining the fluctuations of proteins.\\


\begin{acknowledgments}
This work was supported by
the National Science Foundation Grant CHE-1362500. This work used the Extreme Science and Engineering Discovery Environment (XSEDE), which is supported by National Science Foundation grant number ACI-1053575.
\end{acknowledgments}


\end{document}